\documentclass{elsart}

\usepackage[square,comma]{natbib}
\usepackage{graphicx}
\usepackage{pxfonts}
\usepackage{lineno}

\usepackage{amssymb}
\usepackage{footnote}
\usepackage{multirow}
\usepackage{varwidth}
\usepackage{array}
\usepackage{epstopdf}
\usepackage{url}

\journal{}

\begin{document}
\thispagestyle{empty}
\begin{Large}
\textbf{DEUTSCHES ELEKTRONEN-SYNCHROTRON}

\textbf{\large{Ein Forschungszentrum der
Helmholtz-Gemeinschaft}\\}
\end{Large}

DESY 16-194

October 2016

\begin{eqnarray}
\nonumber &&\cr \nonumber && \cr \nonumber &&\cr
\end{eqnarray}
\begin{eqnarray}
\nonumber
\end{eqnarray}
\begin{center}
\begin{Large}
\textbf{On the Coupling of Fields and Particles in Accelerator and  Plasma Physics}
\end{Large}
\begin{eqnarray}
\nonumber &&\cr \nonumber && \cr
\end{eqnarray}

\begin{large}
Gianluca Geloni,
\end{large}
\textsl{\\European XFEL GmbH, Hamburg}
\begin{large}

Vitali Kocharyan and Evgeni Saldin
\end{large}
\textsl{\\Deutsches Elektronen-Synchrotron DESY, Hamburg}
\begin{eqnarray}
\nonumber
\end{eqnarray}
\begin{eqnarray}
\nonumber
\end{eqnarray}
ISSN 0418-9833
\begin{eqnarray}
\nonumber
\end{eqnarray}
\begin{large}
\textbf{NOTKESTRASSE 85 - 22607 HAMBURG}
\end{large}
\end{center}
\clearpage
\newpage

\begin{frontmatter}



\title{On the Coupling of Fields and Particles in Accelerator and  Plasma Physics}


\author[XFEL]{Gianluca Geloni,}
\author[DESY]{Vitali Kocharyan,}
\author[DESY]{Evgeni Saldin}
\address[XFEL]{European XFEL GmbH, Hamburg, Germany}
\address[DESY]{Deutsches Elektronen-Synchrotron (DESY), Hamburg, Germany}

\begin{abstract}
In accelerator and plasma physics it is generally accepted that there is no need to solve the dynamical equations for particles motion in manifestly covariant form, that is by using the coordinate-independent proper time to parameterize particle world-lines in space-time. In other words, in order to describe the dynamical processes in the laboratory frame  there is no need to use the laws of relativistic kinematics. It is sufficient to take into account the relativistic dependence of the particles momentum on the velocity in the second Newton's law. Therefore, the coupling of fields and particles is based, on the one hand, on the use of result from particle dynamics treated according to Newton's laws in terms of the relativistic three-momentum and, on the other hand, on the use of Maxwell's equations in standard form. In previous papers we argued that this is a misconception. The purpose of this paper is to describe in detail how to calculate the coupling between fields and particles in a correct way and  how to develop a new algorithm for a particle tracking code in agreement with the use of Maxwell's equations in their standard form. Advanced textbooks on classical electrodynamics correctly tell us that Maxwell's equations in standard form in the laboratory frame and charged particles are coupled by introducing particles trajectories as projections of particles world-lines onto coordinates  of the laboratory frame and by subsequently using the laboratory time to parameterize the trajectory curves. For the first time we showed a difference between  conventional and covariant particle tracking results in the laboratory frame. This essential point has never received attention in the physical community. Only the solution of the dynamical equations in covariant form gives the correct coupling between field equations in standard form and  particles trajectories in the laboratory frame. We conclude that previous theoretical and simulation results in accelerator and plasma physics should be reexamined in the light of the pointed difference between conventional  and covariant particle tracking.

\end{abstract}

%
%

%
\end{frontmatter}



\section{ Introduction }

The coupling between charged particles and electromagnetic fields is often described with the help of Maxwell-Lorentz equations in the laboratory frame, which we assume inertial. The full system of equations reads:

\begin{eqnarray}
&& \vec{\nabla}\cdot \vec{E} = 4 \pi \rho~, \cr && \vec{\nabla}\cdot
\vec{B} = 0~ , \cr && \vec{\nabla}\times \vec{E} =
-\frac{1}{c}\frac{\partial \vec{B}}{ \partial t}~,\cr &&
\vec{\nabla}\times \vec{B} = \frac{4\pi}{c}
\vec{j}+\frac{1}{c}\frac{\partial \vec{E}}{\partial t}~ \label{Max}
\end{eqnarray}
together with

\begin{eqnarray}
&& \frac{d\vec{p}_n}{dt} = e_n\left(\vec{E} + \frac{\vec{v}_n}{c}\times \vec{B}\right) ~,\cr &&
\vec{p}_n = m_n\vec{v}_n\left(1 - \frac{v_n^2}{c^2}\right)^{-1/2}~ ,\label{N}
\end{eqnarray}
where the last equations describe the Lorentz force acting on the particles, while the first four equations are the Maxwell's equations. Charge density $\rho$ and current density $\vec{j}$ are given by

\begin{eqnarray}
&&\rho(\vec{x}, t) = \sum_{n} e_n\delta(\vec{x} - \vec{q}_n(t)) ~,\cr &&
\vec{j}(\vec{x},t) = \sum_{n} e_n\vec{v}_n(t)\delta(\vec{x} - \vec{q}_n(t))~. \label{CD}
\end{eqnarray}
Here $\delta(\vec{x} - \vec{q}_n(t))$ is the three-dimensional delta function, $m_n, e_n, \vec{q}_n(t)$, and $\vec{v}_n =   d\vec{q}_n(t)/dt$ denote rest mass, charge, position, and velocity of the $n$th particle, respectively. Finally, the time $t$ is recorded by using clocks in rest relatively to the laboratory frame and synchronized by light-signals.

This coupling of Maxwell's equations and Newton's equations through the Lorentz force is widely used, commonly accepted in accelerator and plasma physics and, in particular, in analytical and numerical calculations of synchrotron and cyclotron radiation (see e.g. \cite{DU,Eck}). In order to evaluate  radiation fields arising from external sources specified by Eq. (\ref{CD}) we need to know, for each particle, the velocity $\vec{v}_n$ and the position $\vec{q}_n$ as a function of the laboratory time $t$. The relativistic motion of each particle in the laboratory frame is described, instead, by Eq. (\ref{N}). In other words, it is generally accepted that, in order to describe the dynamical evolution of a relativistic beam of particles there is no need to use the laws of relativistic kinematics: it is sufficient to take into account the relativistic dependence of the particles momentum on the velocity.

In our previous publications \cite{OURS1,OURS2,OURS3} we argued that this way of coupling fields and particles is misconception, which actually led to a strong qualitative disagreement between theory and experiments. In this paper we present details on how to perform correct calculations pertaining the coupled system of charges and radiation and on how to develop a new algorithm for a particle tracking code. In particular, we will point out that only the solution of the dynamical equations in covariant form gives the correct coupling between the field equations in their usual standard form and the trajectories of particles in the laboratory frame. In other words, Maxwell's equations in their usual standard form in the laboratory frame and the motion of charged particles are coupled by describing the particles trajectories in the laboratory frame with a projection of their world-lines onto the coordinates of the lab frame and by using the laboratory frame time in order to re-parameterize the trajectory curves. In an equivalent, but more mathematical wording, we wish to find the trajectory of any particle in the laboratory frame, $\vec{q}(t)$, consistently with Maxwell's equations in standard form Eq. (\ref{Max}). In order to do so, there is a need to solve the dynamical equation of motion in manifestly covariant form by using the coordinate-independent proper time $\tau$ to parameterize the particle world line $x_\mu(\tau)$:

\begin{eqnarray}
&& m\frac{d^2 x^{\mu}}{d\tau^2} = e F^{\mu\nu}\frac{dx_{\nu}}{d\tau}~ ,\label{CO}
\end{eqnarray}

where $F^{\mu\nu}$ are the components of the electromagnetic field tensor.

Note that the dynamical evolution in the laboratory frame described by Eq. (\ref{N}) is based on the use of the laboratory frame time $t$ as independent variable. In this case, the trajectory $\vec{q}(t)$ can be seen, from the laboratory frame view, as the result of successive Galilean boosts. This evolution of particles in terms of Galilean boosts poses a problem when the particles motion needs to be coupled with Maxwell's equations. In fact, the d'Alembertian, which enters in basic equations of the electromagnetism is not a Galilean invariant.

In contrast, the trajectory $\vec{q}_{cov}(t)$, which is found by using the manifestly covariant dynamical equation, Eq. (\ref{CO}) can be viewed from the laboratory frame as the result of successive Lorentz boosts. In section \ref{due} we will demonstrate in detail for a particular example case the difference between  conventional particle trajectory $\vec{q}(t)$ calculated by solving Eq. (\ref{N}) and  the covariant particle trajectory $\vec{q}_{cov}(t)$ calculated by projecting the world-line $x(\tau)$ onto the laboratory frame coordinates. This essential difference has never received attention in the physical community.

We stress that the  statement made above, that there is a difference between the two trajectories  $\vec{q}(t)$ and $\vec{q}_{cov}(t)$ does not mean that the conventional trajectory  $\vec{q}(t)$ is incorrect. Within the framework of dynamics only, both trajectories describe correctly the same physical reality. Different expressions for the particle trajectories are different only because they are based on the use of different clock synchronization conventions. Whenever we have a theory containing an arbitrary convention, we should examine what parts of the theory depend on the choice of that convention and what parts do not. We may call the former part convention-dependent, and the latter convention-invariant. Clearly, physically meaningful  results must be convention-invariant.  We state that the difference between the two trajectories  $\vec{q}(t)$ and $\vec{q}_{cov}(t)$ is convention-dependent and has no direct physical meaning. Once more, different expressions for particle trajectory in a single (e.g. the laboratory) reference system arise from the use of different  synchronization conventions.

Different types of clocks synchronization obviously provide different time coordinates that describe the same reality: in particular, we cannot give any experimental method by which simultaneity between two events in different places can be ascertained. In other words, the determination of simultaneous events implies the choice of a convention. In a similar fashion, in order to measure the speed of a particle, one first has to synchronize the clocks that measure the time interval as the particle travels between two given points in space. Therefore it can be said that, consistently with the conventionality of simultaneity, the value of the particle speed is also  a matter of convention and has no definite objective meaning. Let us consider, for instance, the case of a circular or helical motion in a constant magnetic field: the particle speed is convention-dependent.  In contrast to this, the radius of rotation has a direct objective meaning, and does not depend on the choice of clock synchronization.

Note that advanced textbooks on classical electrodynamics correctly tell us that Maxwell's equations  and charged particles are coupled in a covariant manner, see e.g. \cite{J, SCH}. The charge and current densities in Eq. (\ref{CD}) can, in fact, be written as a four-current vector with coordinates $j_\mu(x)$ by introducing the charge 4-vector coordinate $x_{\mu}(\tau)$ as a function of the charge proper time $\tau$ and integrating over the proper time with an appropriate additional delta function. This leads to

\begin{eqnarray}
&& j_{\mu}(x) = ec\int d\tau  ~u_{\mu}(\tau) \delta^4(x - x(\tau)) ~ ,\label{SO}
\end{eqnarray}
where the charge 4-velocity $u_{\mu}(\tau)$ and the 4-vector coordinates $x_{\mu}(\tau)$ are solutions of the covariant equation of motion Eq. (\ref{CO}). Integration over the proper time of $\tau$ leads to

\begin{eqnarray}
&& j_{\mu}(\vec{x},t) = eu_{\mu}(t)\delta^3(\vec{x} - \vec{q}_{cov}(t)) ~ ,\label{FC}
\end{eqnarray}
where $\vec{q}_{cov}(t)$ is what we are looking for: it is the particle trajectory in the laboratory frame found as the result of the projection of the particle world-line onto the coordinates of the laboratory frame. As said before, $\vec{q}_{cov}(t)$ is consistent with Maxwell's equations in standard form Eq. (\ref{Max}). Unfortunately, the difference between conventional particle trajectory  and covariant particle trajectory seems to have been almost entirely overlooked by many physicists (with the exception of Lorentz
\footnote{There are two satisfying ways of coupling fields and particles. The first, Einstein's way, consists in using relativistic kinematics for the description of the particle beam evolution and Maxwell's equations in standard form. The second, Lorentz's way, consists in using the conventional trajectories $\vec{q}(t)$ and in a reformulation of Maxwell's equations in the absolute time synchronization convention.} \cite{LO})  who developed results of the theory of radiation from charged particles by using Maxwell's equations in standard form Eq.(\ref{Max}) and  $\vec{q}(t)$ instead of $\vec{q}_{cov}(t)$.

Curiously,  even  first order optical phenomena like the aberration of light cannot be explained on the basis of the conventional coupling of fields and particles (Eq.(\ref{Max}) and  $\vec{q}(t)$), which gives a result in contradiction with experiments. We conclude that previous theoretical results in radiation theory and, in particular, in synchrotron radiation theory should be reexamined in the light of the pointed difference between conventional  and covariant particle trajectories.

\section{\label{due} Explicit example of the difference between the trajectories $\vec{q}(t)$ and $\vec{q}_{cov}(t)$ }

\begin{figure}
\begin{center}
\includegraphics[width=0.8\textwidth]{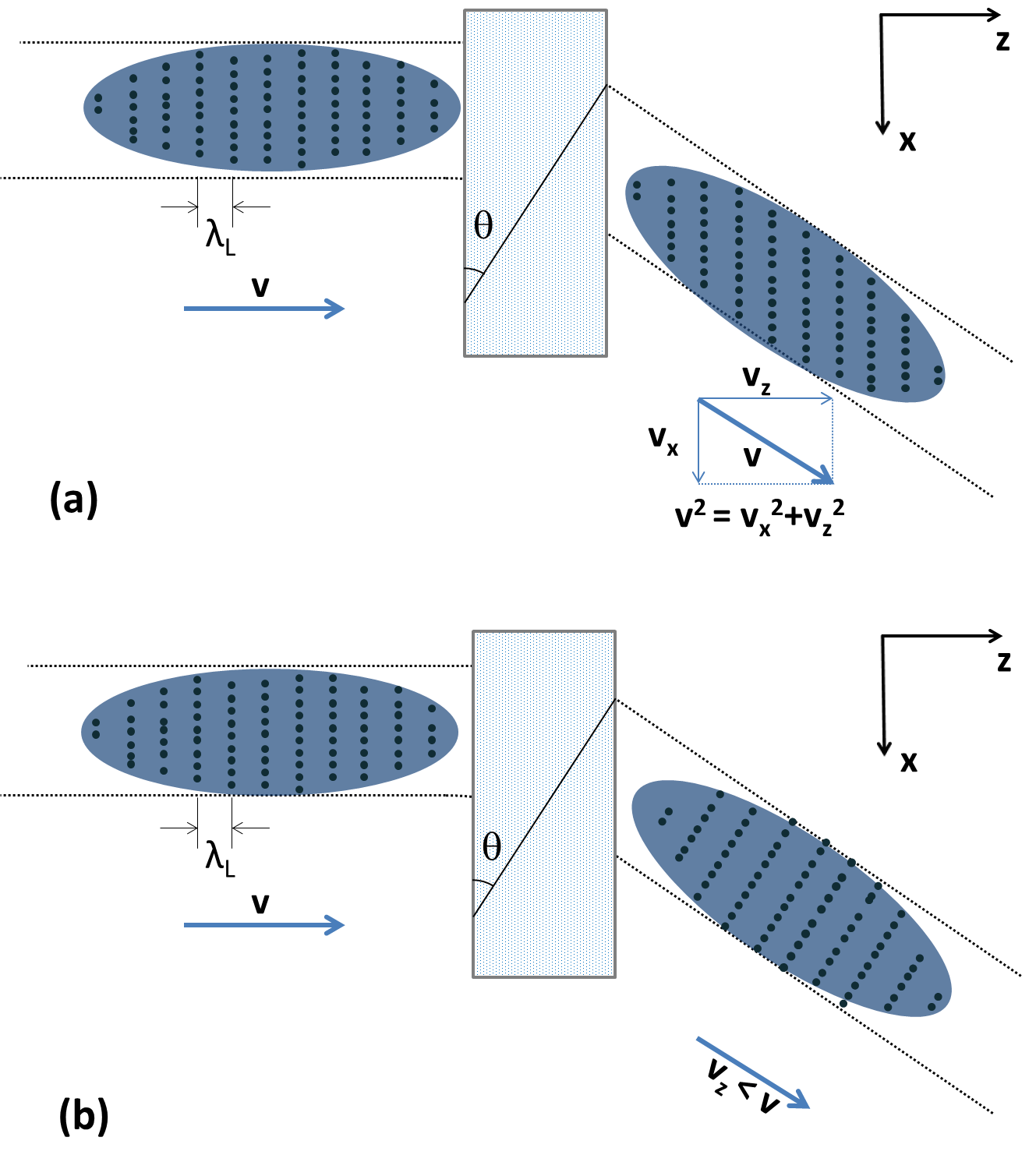}
\end{center}
\caption{A microbunched, ultrarelativistic electron beam undergoes a kick of an angle $\theta$. Different conventional choices give different results for the electron beam evolution. (a) Conventional particle tracking can be described in terms of a Galileo transform:  simultaneity is preserved; the microbunching direction remains unvaried as well as the speed of the electrons, $v$.  (b) Another conventional choice consists in using Lorentz boosts to describe the particles evolution: simultaneity is not preserved in this case; the microbunching direction is tilted and the velocity of the electrons changes from $v$ to $v_z<v$.} \label{uno}
\end{figure}

In this section we present a study  of an experimental setup  for illustrating the difference between conventional and covariant trajectories. We have chosen this particular example because it is relatively simple, although it can stand as a basic case that can be generalized to describe all synchrotron radiation phenomena. This study also has practical applications, since it provides a method to correctly analyze the effect of trajectory errors on the amplification process of X-ray Free-Electron Lasers (XFELs). Let us consider the simple case when a microbunched ultrarelativistic electron beam is kicked by a weak dipole field before entering a downstream undulator and study the process of emission of coherent undulator radiation. Fig. \ref{uno} shows a schematics of a microbunched electron beam undergoing a kick.  Conventional particle tracking (based on Eq. (\ref{N})) states that after the beam is kicked there is a trajectory change, while the orientation of the microbunching phase front remains as before.  In other words, the kick results in a difference between directions of the electron motion and the normal to the phase front (see Fig. \ref{uno}(a)). According to conventional particle tracking, a kick along the $x$ direction is equivalent to a Galilean coordinate transformation as $x' = x - v_x t$.  This transformation is completed with the invariance of simultaneity; in other words, if two electrons arrive simultaneously at a certain position $z$ down the beam, i.e. $\Delta t = 0$, then after the transformation above the same two electrons reach position $z'=z$  once more simultaneously, i.e. $\Delta t'=0$. The absolute character of temporal simultaneity between two events is a consequence of the identity $t' = t$.  As a result of the kick, the transformation of time and spatial coordinates of any event has the form of a Galileo boost rather than a Lorentz boost.

In conventional particle tracking, the simultaneity along the $x$ direction has an absolute character, meaning that it is independent of the kick. According to the theory of relativity, however, we can establish another criterion for the simultaneity of events, which is based on the invariance of the speed of light. It is immediately understood that, as a result of  the motion of electrons along the wavefront after the kick, the simultaneity of different events is no longer absolute, i.e. independent of the kick. This reasoning is in analogy with Einstein's train-embankment thought experiment. There is a reason for arguing that we must have made a mistake. Of course, we did not make a mathematical mistake in the solution of Lorentz force equation Eq. (\ref{N}), but may be we have left something out. As we already mentioned above, this is no mistake with point of view of dynamics. The microbunching wavefront can be considered as a plane of simultaneous events. Establishing simultaneous events is only a matter of convention and the orientation of the microbunching wave vector of an ultrarelativistic electron beam has no definite objective meaning.

In contrast to this, the  direction of emission of coherent radiation from a microbunched electron beam obviously has a direct objective meaning. In standard electrodynamics, coherent radiation is emitted in the direction normal to the microbunching wavefront. Therefore, according to the conventional coupling of fields and particles, when the angular kick exceeds the divergence of the output radiation emission in the direction of the electron beam motion is strongly suppressed. There are two outstanding predictions of conventional theory concerning the beam kicking experimental setup in Fig. \ref{uno}(a). (see e.g. \cite{TA}). The first is that a kicked electron beam coherently radiates towards the wavefront normal i.e. along the $z$ axis. After the kick, the beam velocity components are $(v_x,0, v_z)$, where $v_z = \sqrt{v^2 - v_x^2}$ and $v$ is the beam velocity along  the $z$-axis (i.e. the undulator axis ) upstream of the kicker. The second prediction of the conventional theory is that if a microbunched beam is at perfect undulator resonance without kick, then after the kick  there is a red shift in the resonance wavelength in the radiation direction (i.e. along the $z$-axis). The maximum power of the coherent radiation after the kick is reached when the undulator is detuned to be resonant to the lower longitudinal velocity  after the kick.

According to conventional particle tracking, any transformation of observations in the comoving inertial frame where the particle is instantaneously at rest to the laboratory frame can be made by Galilean transformations. If a relativistic particle is accelerating in the laboratory frame, one can think of successive Galileo boosts that track the motion of the accelerated particle. The usual Galileo rule for adding velocities is used to fix the Galileo boosts tracking that particular particle along its motion. The use of a sequence of Galileo boosts to describe the motion of relativistic particles is puzzling, because the equation of motion Eq. (\ref{N}) is consistent with the principle of relativity. We would like, therefore, to clarify how Galilean transformations can be understood in terms of the theory of relativity. Since the formulation of special relativity, most researchers assume that Lorentz transformations between inertial frames immediately follow from the postulates of the theory of relativity. However, these postulates alone are not sufficient to obtain Lorentz transformations: one additionally needs to synchronize spatially separated moving clocks with the help of light signals. If this done using Einstein's synchronization convention, then Lorentz transformation follow. However, if the same clocks are synchronized following a different synchronization convention, other transformations follow.

In order to get a Galilean transformation, we should synchronize clocks in the laboratory system  with the usual procedure suggested by Einstein and involving light signals. Next, in order to perform measurements in a moving inertial frame (a frame moving with velocity $v_x$ with respect to the laboratory system), it is necessary to synchronize the moving clocks.  This can be done with the help of clocks at rest without light signals simply by adjusting them according to the reading of a clock at rest whenever they fly past it (see e.g. \cite{GU}). In other words, on the basis of Newton law Eq. (\ref{N}), it is assumed that moving clocks are externally synchronized i.e. synchronized with the help of the clocks at rest  in lab frame (absolute time convention). Then, according to the conventional coupling of fields and particles, the sources in Eq. (\ref{CD}) are the result of the transformation of source observations made in the moving frame to the laboratory frame, done with Galilean transformations. Consistently, these conventional sources in the laboratory frame should be coupled to the field equations found by transforming standard Maxwell's equations in the moving frame to the laboratory frame with Galilean transformations as well. Maxwell's equations, however, do not remain form-invariant with respect to Galilean transformation. In fact, the d'Alembertian operator, which enters basic electrodynamical equations is not Galilean invariant.  Its change of form under Galilean transformation can be verified by replacing $\partial/\partial t = \partial/\partial t' - v_x\partial/\partial x'$, $\partial/\partial x = \partial/\partial x'$.  We must conclude that
coupling the electrodynamical equations Eq. (\ref{Max}) to Newton equation with Lorentz force Eq. (\ref{N}) is a misconception.

According to covariant particle tracking, any transformation of observations in the comoving inertial frame where the particle is instantaneously at rest to the laboratory frame can be done by Lorentz transformations. If a relativistic particle is accelerating in the laboratory frame, one can think of successive Lorentz transformations  that track the motion of the accelerated particle. The Einsten's  rule for addition of velocities is used to fix the Lorentz transformations tracking that particular particle along its motion. In order to get a Lorentz transformation,  we must synchronize clocks in the laboratory frame  with the usual Einstein procedure involving light signals, as before. However, in contrast to what has been done before, in order to perform source observations in a moving inertial frame, we use clocks in rest relative to the moving frame and synchronized by light signals.  In this case the d'Alembertian remains form-invariant with respect to Lorentz transformations. We may now summarize of the main conclusion of our discussion: we state that Maxwell's equations in their standard form are consistent with particle trajectories in the laboratory frame described as a sequence of Lorentz boosts tracking a particle.

We now wish to consider in detail an experimental setup with a microbunched electron beam and a kicker, as shown in Fig. \ref{uno}, and present an analysis of how a sequence of Lorentz boosts unfolds, giving rise to the behavior in Fig. \ref{uno}(b). Suppose that an electron beam moves, initially, at ultrarelativistic velocity $v$ parallel to the $z$-axis upstream the kicker, assuming for simplicity that the kick angle $\theta \simeq v_x/c$ is small compared with $1/\gamma$, where $\gamma = 1/\sqrt{1-v^2/c^2}$  is the relativistic factor. This means that we take the limit $\gamma^2v_x^2/c^2 \ll 1$.

Let us consider a composition of Lorentz boosts that track the motion of a relativistic electron accelerated by the kicker field. Let $S$ be the laboratory frame of reference, and $S'$ another inertial frame moving with velocity $\vec{v}$ relative to $S$. Particles upstream of the kicker are at rest with respect to $S'$. By definition, $S'$ is connected to $S$ by the Lorentz boost $L(\vec{v})$ according to $X' = L(\vec{v})X$, where we denote with $X$ a four vector describing an event in a space-time with respect to $S$. We turn our attention to what happens in $S'$, in which particles are at rest and the kicker is running  with velocity $-\vec{v}$ towards them. In $S'$, the moving magnetic kicker produces an electric field. Therefore, in order to describe the kick, we must consider electrons moving in a combination of electric and magnetic fields perpendicular to each other.  It is easy to see that the acceleration in these crossed fields  gives rise to the electron beam velocity specified by $v'_x =\gamma v_x$ parallel to the $x$-axis and  $v'_z = - v(\gamma v_x/c)^2/2$ parallel to the $z$-axis. If  $\gamma^2v_x^2/c^2 \ll 1$, the motion of the kicked beam in the moving reference frame $S'$ is non relativistic. Let now $S"$ be a third inertial frame where particles downstream of the kicker are at rest. In order to relate $S"$ to $S'$ we should perform a composition of boosts along the $z$ axis and along the $x$ axis. However, as is known, the composition of non-collinear Lorentz boosts does not result in another Lorentz boost but rather in a more complicated Lorentz transformation involving a boost and rotation, the Wigner rotation \cite{WI,WI1,WI2}. In our non-relativistic asymptotic we can neglect the fact that the two non-collinear Lorentz boosts do not commute and use the transformation $X"=L(\vec{e}_x v'_x) L(\vec{e}_z v'_z)X' =  L(\vec{e}_z v'_z)L(\vec{e}_x v'_x)X'$ to discuss the beam motion in $S'$ after the kick. Here  $\vec{e}_x$ and $\vec{e}_z$ are unit vectors respectively directed along the $x$ and the $z$ axis. The sequence of Lorentz boosts $L(\vec{e}_x v'_x)L(\vec{e}_z v'_z)L(\vec{e}_z v)$ presents a step-by-step change from $S$ to $S'$ and then to $S"$ according to $X" = L(\vec{e}_x v'_x)L(\vec{e}_z v'_z)L(\vec{e}_z v)X$. For the first two boosts,  velocities are parallel and the addition law is $L(\vec{e}_z v'_z)L(\vec{e}_z v) = L(\vec{e}_z v_z)$. Here
$v_z = v(1 - \theta^2/2)$ and $\theta = v_x/c$. The resulting boost composition can be represented as $X" = L(\vec{e}_x v'_x)L(\vec{e}_z v)X$. As discussed before, this product of two non-collinear boosts is not a boost, but it can be represented as composition of a boost and a three-dimensional rotation: $L(\vec{e}_x v'_x)L(\vec{e}_z v_z)
= R(\theta)L(\vec{n}v_z)$ , where $R(\theta)$ is the matrix of the  rotation of the $S"$ system through an Wigner angle $\theta = v_x/c$ in the $x,z$ plane of the system $S$ and $\vec{n}$ is the unit vector  $\vec{n} = \vec{e}_x\theta + \vec{e}_z (1-\theta^2/2)$ (up to the fourth order in $\theta$. We note that the interpretation of the Wigner rotation in the laboratory frame of reference requires a certain care \cite{MAL,RI1, ST}. We also note that we discuss particle tracking under the approximation $v_x^2/c^2 \ll \gamma^2v_x^2/c^2 \ll 1$. However, even in this simple example we are able to demonstrate  the difference between conventional and covariant particle trajectories. The microbunching orientation is readjusted along the new direction  of the electron beam and the speed of electrons decreases from $v$ to $v_z$, see Fig. \ref{uno}(b). The result is at odds with the prediction from  conventional particle tracking, see Fig. \ref{uno}(a).

We can look at this result of covariant particle tracking in the following way: if the velocity of a modulated electron beam is close to the velocity of light, Lorentz transformations work out in such a way that the rotation angle of the microbunching wave vector coincides with the angle of rotation of the velocity. In this case,  the length of the wave vector along the direction of the bunch motion is  Lorentz invariant. This is plausible if one keeps in mind that the wave vector of a laser pulse behaves precisely in the same way: during the motion along a curvilinear trajectory, the wave vector of the radiation is always aligned with the direction of motion of the laser pulse. It follows from the previous reasoning that in the large momentum (or zero mass) limit, whatever we know about the kinematics of a laser pulse can immediately be applied to an ultrarelativistic modulated electron bunch.

There are physical consequences of this new particle tracking rule. Let us see what happens if the kicked electron beam enters a downstream undulator. There is a strong qualitative disagreement between predictions of covariant and conventional coupling of fields and particles concerning this "beam kicking" experimental setup. First, the kicked electron beam coherently radiates towards the microbunching wavefront normal i.e., in covariant coupling, along the kicked direction.  Second, if the microbunched beam is at perfect undulator resonance without kick, then after the kick, according to Fig. \ref{uno}(b),  there is red shift in the resonance wavelength in the kicked direction. We have the result that the maximum power of the coherent radiation in the kicked direction is reached when the undulator is detuned to be resonant to the lower velocity  after the kick.

We now wish to consider an experiment whose results can only be explained on the basis of our covariant coupling of fields and particles. We refer to the recent "beam splitting" experiment at the LCLS \cite{NUHN}. It apparently demonstrated that after a microbunched electron beam is kicked on a large angle compared to the divergence of the FEL radiation  \footnote{ The tuning limit of the deflection angle  was set by the beamline aperture to $\sim$ 5 rms of the FEL radiation divergence, see Fig. 14 in \cite{NUHN}}, the microbunching wavefront is readjusted along the new direction of motion of the kicked beam. Therefore, coherent radiation from the undulator placed after the kicker is emitted along the kicked direction practically without suppression. The results of the "beam splitting" experiment at the LCLS, demonstrated that even the direction of emission of coherent undulator radiation is beyond the predictive power of the conventional theory.

In the framework of the conventional theory, there is also a second outstanding puzzle concerning the beam splitting experiment at the LCLS. In accordance with conventional coupling of fields and particles, if the microbunched beam is at perfect (undulator) resonance without kick, then after the kick the same microbunched beam  must be at  perfect resonance in the kicked direction. This is plausible, if one keeps in mind that, after the kick, the particles have the same velocity  and emit radiation in the kicked direction owing to the Doppler effect. However, experimental results clearly show that there is a red shift in the resonance wavelength in the kicked direction.  The maximum power of the coherent radiation after the kick is reached when  undulator is detuned to be resonant to the lower longitudinal velocity after the kick \cite{NUHN}.

Here we have shown that manifestly covariant coupling of fields and particles predict a surprising effect, in complete contrast to the conventional treatment. Namely, according to covariant particle tracking, the plane of simultaneity (i.e. the microbunching phase front) orientation  in the ultrarelativistic asymptotic is always perpendicular to  the beam velocity. This effect allows for the production of coherent undulator radiation from the  modulated electron beam in the kicked direction. It is necessary to mention that in the case of the beam splitting experiment at the LCLS we deal indeed with an ultra relativistic  electron beam ($c-v\simeq 10^{-8}c )$, and with  a transverse velocity after the kick, which is very much smaller than speed of light ($(v_x/c)^2 \ll 10^{-8}$), so that the theoretical studies presented above yields a correct quantitative description of the beam splitting experiment at the LCLS and, in particular, of the red shift in the resonance wavelength in the kicked direction.

We have described a simple experiment that directly illustrates the difference between
conventional and covariant trajectories. Furthermore, we have restricted our discussion to a simple case and have assumed that kick angle is sufficiently small, $(\gamma\theta)^2\ll 1$. Surprisingly, the essential aspects of X-ray FEL behavior can be discussed within the presented framework. The performance of an XFEL can be reduced if any of a number of undulator  parameters, deviates from its optimum value. Analysis of the effect of trajectory errors on the XFEL amplification process showed that it becomes more and more important at shorter wavelength. Previous numerical studies of this critical aspect in the design of an XFEL source, however, was based on the conventional coupling of fields and particles i.e. on an incorrect kinematical model \cite{TA}. Therefore, the tolerances  predicted are more stringent than they need be. This can be considered one of the reasons for the exceptional progress in the Angstrom-wavelength  XFEL developments over the last  decade.

\section{Discussion and Conclusions}

The experiment mentioned in the preceding section may be regarded as a decisive experimental verification of predictions of manifestly covariant coupling fields and particles. However, it is not necessary at all to perform complicated experiments. Instead, one could thoroughly analyze the results of known experiments. We point out  that on the basis of conventional coupling of fields and particles even the effect of light aberration cannot be explained.

To see this, suppose that we have a plane full of sources, all oscillating together, with their motion on the plane, and all having the same amplitude and phase. If we let the plane of charges be the $xy$-plane,  then a simple Huygens' construction shows that this emitters will radiate plane wavefronts along the $z$-axis.  When a kick is introduced, emitter moves at constant speed $v_x$ along the  plane of sources.  After the kick along $x$-axis, Cartesian coordinates transform according to $x' = x - v_xt$, $y' = y$, $z' = z$. This transformation is completed by the invariance in simultaneity  $\Delta t = \Delta t'$. The absolute character of the simultaneity of two events is a consequence of the absolute concept of time $t' = t$. As a result, a kick transformation of time and spatial coordinates $t,x,y,z$ of any event has the form of a Galilean transformation. We come to the situation when there is a motion of elementary sources along the plane of simultaneity. Lorentz was able to show that charge dynamics under the convention of absolute simultaneity is in contradiction with  Maxwell's equations in standard form and found a satisfying, pre-relativistic way of coupling dynamics and electrodynamics in this situation when the solution of the dynamical problem was performed in the absolute-time world-picture. Lorentz' pre-relativistic  way consists in a sort of "translation" of Maxwell's electrodynamics to the absolute time world-picture \cite{LO}.

Let us consider the  "translation" of the d'Alembertian to the absolute-time world-picture. After properly transforming the d'Alembertian we can see that the inhomogeneous wave equation for the electric field in the laboratory frame after a Galileo boost has nearly but not quite  the usual, standard form that takes when there is no common, uniform translation of charges in the transverse direction with velocity $v_x$. The main difference consists in the "interference" term $\partial^2/\partial t\partial x$ which arises when applying our Galileo boost. The discussion can be simplified by the use of a mathematical trick, without the direct solution of the modified wave equation. Lorentz found that the solution of the electrodynamical problem in the absolute-time convention can be obtained with minimal efforts by formally desynchronizing the absolute time (which Lorentz called the "true" time) $t$ to the "local" time $t' = t - x v_x/c^2$ and using $t'$ without changing the d'Alembertian \cite {LO}. It is immediately seen by direct calculations that a shift of time is what is needed in order to the eliminate interference term. The effect of this time transformation is just a dislocation in the  timing of processes. This transformation  has the effect of rotating of the plane of simultaneity on the angle  $v_x/c$ in the first order approximation. In that case, due to motion of the emitter, the elementary sources along it produce a spatial phase modulation  (chirp) $\omega x v_x/c^2$, where $\omega$ is the oscillation frequency. As a consequence of this linear phase chirp, the wavefront propagates at speed $c$ with the aberration angle $\theta = v_x/c$ from the $z$-direction.

The effect of light aberration can also be explained on the basis of relativistic kinematics, when the dynamics evolution is treated under the Einstein's time order. On the one hand, it is well known that the wave equation remains  form-invariant with respect to Lorentz transformations. On the other hand, if we make a Lorentz boost in the $x$-direction to describe the uniform translation along $x$-axis in the laboratory frame, we automatically introduce the "local" time  $t' = t - xv_x/c^2$ and the effect of this transformation is just a rotation of the wave front. In other words, in the first order in $v_x/c$, the Galilean transformation described above, completed by the introduction of the "local" time is mathematically equivalent to the Lorentz transformation just described here: it does not matter which convention and hence transformation or "translation"  is used to describe the same reality. We note that even in the non-relativistic limit, when we can neglect second order corrections in $v_x/c$, which are intrinsically relativistic, Lorentz and Galileo transformations are different. The difference is in the term $x v_x/c^2$ in the Lorentz transformation for time, which is a first order correction. Yet we underline that this term is only conventional and has no direct physical meaning. In other word, differences that arise between Galilean and Lorentz transformations in the non-relativistic limit are only to be ascribed to the use of different synchronization conventions.

For the particular experiment of Fig. \ref{uno} we showed that one of the immediate consequences of the new particle tracking rule is the occurrence  of a red shift of the resonance wavelength, which arises in the kicked direction. The maximum of the coherent undulator radiation after the kick is reached when the undulator is detuned to be resonant to the lower velocity $v_z$,  Fig. \ref{uno}(b). Now we can imagine a similar experimental setup with only a single electron: the presence of a red shift in coherent undulator radiation would still be there in the case of spontaneous emission from a single electron. This red shift in the kicked direction cannot be explained in the framework of conventional synchrotron radiation theory. Clearly, the conventional  theory predicts a zero red shift for a fundamental reason related to the Doppler effect. In fact, before and after the kick, the electron has the same speed, Fig. \ref{uno}(a).  Therefore, in terms of conventional synchrotron radiation theory, the particle emits spontaneous undulator radiation without red shift in the kicked direction  \cite{K,K1}.

There is another interesting spontaneous emission  problem where a correction of conventional synchrotron radiation theory is required. The presence of red shift in undulator radiation automatically implies the same red shift also in the case of conventional synchrotron-cyclotron radiation theory. The spectral and angular behavior of synchrotron-cyclotron radiation emitted by an electron moving in a constant magnetic field, and having an ultrarelativistic velocity component perpendicular to it is described by well-known analytical formulas. At present,  relativistic synchrotron-cyclotron radiation results are considered textbook examples and do not require a detailed description.
These results should also be reexamined in the light of the pointed difference between conventional and covariant particle tracking.

\section{Acknowledgements}

We greatly thank Martin Dohlus, Oleg Gorobtsov, Robin Santra, Svitozar Serkez, Joachim Stohr,   and Igor Zagorodnov for useful discussions, Franz-Joseph Decker, Zhirong Huang, James MacArthur and Heinz-Dieter Nuhn, for discussion  about beam splitting experiment at the LCLS.

\end{document}